# Applying weighted PageRank to author citation networks

Ying Ding

School of Library and Information Science, Indiana University,

1320 East 10th Street, Herman B Wells Library, LI025, Bloomington, IN 47405, USA

Tel: (812) 855 5388, Fax: (812) 855 6166, Email: dingying@indiana.edu

## Abstract

This paper aims to identify whether different weighted PageRank algorithms can be applied to author citation networks to measure the popularity and prestige of a scholar from a citation perspective. Information Retrieval (IR) was selected as a test field and data from 1956-2008 were collected from Web of Science (WOS). Weighted PageRank with citation and publication as weighted vectors were calculated on author citation networks. The results indicate that both popularity rank and prestige rank were highly correlated with the weighted PageRank. Principal Component Analysis (PCA) was conducted to detect relationships among these different measures. For capturing prize winners within the IR field, prestige rank outperformed all the other measures.

## Introduction

The word "*popularity*" is derived from the Latin word popularis[1] and its meaning evolved from "*belonging to the people*" to "*being beloved by the people*." Prestige suggests "*important popularity*," which may be described as "*reputation*", "*esteem*", "*high standing among others*", or "*dazzling influence*."[2] Bibliometrically, the popularity of a researcher can be measured by the number of citations he has accumulated, and his prestige can be calibrated by the number of citations from highly cited publications (see Figure 1, Ding & Cronin, 2010 forthcoming). Researchers can be popular but not necessarily prestigious or vice versa. For example, an author of an article introducing trendy topics in one field can be cited by many young researchers who are relatively new to the field, but may not be cited by domain experts. In contrast, a researcher of a seminal paper introducing innovative methods may be highly appreciated by domain experts, but not laymen. Prestige, therefore, indicates important popularity.

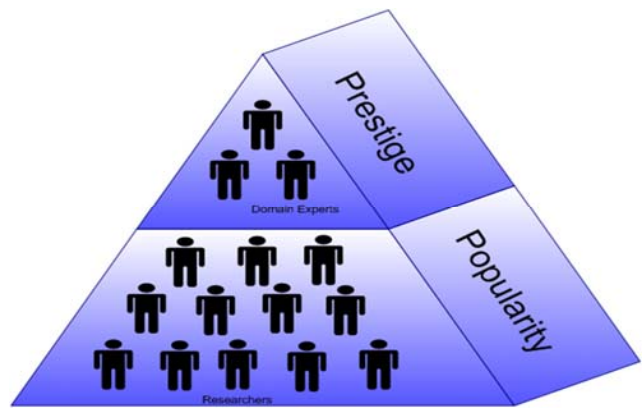

Figure 1. Prestige (cited by highly cited papers) and Popularity (cited by normal papers).

Internet search engines need to distinguish between websites linked by normal websites from those linked by important websites (e.g., Google, Yahoo, IBM, Microsoft websites). They face the same issue when identifying prestigious websites. PageRank, invented by Sergey Brin and Lawrence Page, assumes that the importance of any website can be judged by the websites that link to it, which was coincidently motivated by citation analysis. Brin and Page (1998) emphasized the important link between PageRank and citation analysis, stating that "*Academic citation literature has been applied to the web, largely by counting citations or backlinks to a given page. This gives some approximation of a page's importance or quality. PageRank extends this idea by not counting links from all pages equally, and by normalizing by the number of links on a page*" (p. 109). Google uses PageRank algorithm to rank websites by not only counting the number of hyperlinks a website has, but also the number of hyperlinks pointed to by important websites.

In citation analysis, the number of citations reflects the impact of a scientific publication. This measurement does not differentiate the importance of the citing papers: a citation coming from an obscure paper has the same weight as one from a groundbreaking, highly cited work (Maslov & Redner, 2008). Pinski and Narin (1976) were these first scholars to note the difference between popularity and prestige in the bibliometric area. They proposed using the eigenvector of a journal citation matrix (i.e., similar to PageRank) corresponding to the principal eigenvalue to represent journal prestige. Bollen, Rodriguez and Van de Sompel (2006) defined journal prestige and popularity, and developed a weighted PageRank algorithm to measure them. They defined popular journals as those journals cited frequently by journals with little prestige, and prestigious journals as those journals cited by highly prestigious journals. Their definitions are recursive. Recently, Ding and Cronin (2010 forthcoming) extended this approach to authors and applied weighted citation counting methods to calculate researcher prestige in the field of information retrieval. They defined the popularity of a researcher as "*the number of times he is cited (endorsed) in total, and prestige as the number of times he is cited by highly cited papers.*"

The main concept behind this prestige measure is to use simple citation counting but give more weights to highly cited papers.

Since scholarly activities form complex networks where authors, journals, and papers are connected via citing/being cited or co-occurred, the network topology can significantly influence the impact of an author, journal, or paper. The recent development of large-scale networks and the success of PageRank demonstrate the influence of the network topology on scholarly data analysis. PageRank or weighted PageRank have performed well in representing the prestige of journals (Bollen, Rodriguez, & Van de Somple, 2006; Falagas, Kouranos, Arencibia-Jorge, & Karageorgopoulos, 2008), while only a few researchers have applied this concept to authors (Radicchi, Fortunato, Makines & Vespignani, 2009; Zyczkowski, 2010). This paper is built upon the effort of Ding and Cronin (2010 forthcoming) and clearly addresses this need by testing whether PageRank or weighted PageRank algorithms applied in author citation networks can be used to denote the popularity and prestige of scholars.

Information retrieval (IR) field was chosen as the test field and 15,370 papers with 341,871 citations covering 1956-2008 from WOS were collected. Weighted PageRank algorithms were applied in author citation networks and were compared with related measures such as citation counts, the h-index, and popularity and prestige ranks from Ding and Cronin (2010 forthcoming). Measures were also tested by comparing the number of award winners included in their top 5, 10, 20, and 50 lists. This paper is organized as follows. Section 2 discusses related work on popularity and prestige in the bibliometric setting; Section 3 explains the methods used in this study and proposes weighted PageRank algorithms; Section 4 presents and compares original and weighted PageRanks with popularity and prestige ranks from Ding and Cronin (2010, forthcoming), tests the correlations of different measures, and evaluates the different coverage of IR field prize winners for these measures. Finally, Section 5 draws the conclusion and addresses areas of future research.

## Related Work

In bibliometrics, the impact factor is widely used to measure journal prestige (Garfield, 1999; Bordons, Fernandez, & Gomez, 2002; Harter & Nisonger, 1997; Nederhof, Luwel, & Moed, 2001). This same principle has been extended to measure the impact of web spaces (Smith, 1999; Thelwall, 2001). The h-index is used to assess the performance of researchers (Hirsch, 2005; Cronin & Meho, 2006), and different h-index variations have been proposed and studied in recent years (Jin, Liang, Rousseau, & Egghe, 2007; Sidiropoulos, Katsaros, & Manolopoulos, 2007). Redner (1998) measured the popularity of scientific papers based on the distribution of citations. However, these previous measures simply use citation counts, and do not distinguish citations coming from different authors, nor consider the network features of citing behavior.

Network features of citations can be important to differentiate the impact of author, journal and paper. Bollen, Rodriguez and Van de Sompel (2006) proposed the notion of popular journals and prestigious journals. They used a weighted PageRank algorithm to measure journal prestige. They argued that the ISI Impact Factor (ISI IF) is a metric of popularity rather than of prestige. They found significant discrepancies between PageRank and ISI IF in terms of measuring the journal status at their top ranks. Franceschet (2010) conducted a thorough bibliometric analysis to identify the difference between popularity and prestige of journals in science and social science. He used five-year impact factor to represent the popularity of journals and the eigenfactor metric for the prestige of journals. He found various diverging ranks in measuring the popularity and prestige of science and social science journals.

PageRank provides a computationally simple and effective way to identify important nodes in the connected graphs (Maslov & Redner, 2008; Falagas, Kouranos, Arencibia-Jorge, & Karageorgopoulos, 2008). The weighted PageRank is a further extension of PageRank that adds weights to different parts of the PageRank formula. Radicchi, Fortunato, Makines and Vespignani (2009) proposed a weighted PageRank algorithm on a directed weighted author citation network for ranking scientists by considering the diffusion of their scientific credits. Zyczkowski (2010) proposed weighting factors to measure author's impact by normalizing eigenvectors with the largest absolute eigenvalue for author citation networks. Xing and Ghorbani (2004) added weights to the links based on their reference pages, differentiating between inbound and outbound link weights. Their simulation results showed that weighted PageRank outperforms the original PageRank in terms of returning larger number of relevant pages to a given query. Aktas, Nacar, and Menczer (2006) proposed a weighted PageRank algorithm based on user profiles by adding weights to certain Internet domains preferred by users. Yu, Li, and Liu (2004) added a temporal dimension to the PageRank formula by weighting each citation by date. Walker, Xie, Yan, and Maslov (2007) proposed a CiteRank algorithm by introducing two parameters: the inverse of the average citation depth and the time constant which is biased toward more recent publications. Liu, Bollen, Nelson, and Sompel (2005) defined AuthorRank as a modification of PageRank that considers link weights among the co-authorship links.

## Methodology

**Data Collection**

Information retrieval (IR) was selected as the testing field. Papers and their citations were collected from Web of Science (WOS) published between 1956 and 2008. (for more details, see

Ding & Cronin, 2010 forthcoming). In total, 15,370 papers with 341,871 citations were collected. The 15,370 papers did not include book reviews, software or database reviews. The citation records contained the first author, year, source, volume, and page number. The whole dataset was divided into four sets based on the time span: Phase 1 (1956-1980), Phase 2 (1981-1990), Phase 3 (1991-2000) and Phase 4 (2001-2008). Figure 2 shows the publication distribution among these four phases.

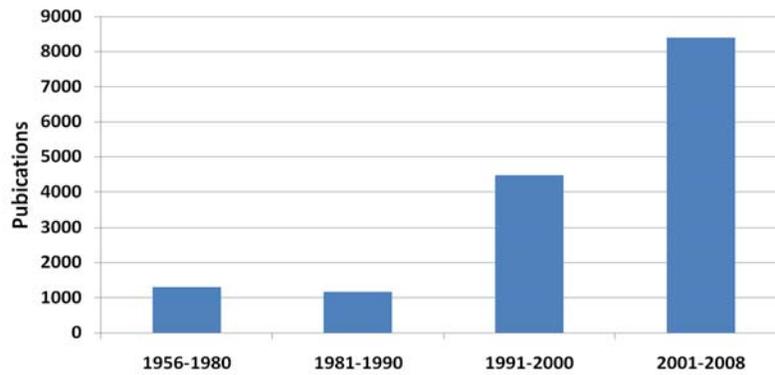

Figure 2. IR publications

**PageRank and Weighted PageRank**

The PageRank algorithm can be defined by the following equation

$$PR(p_i) = \frac{1-d}{N} + d \sum_{p_j \in M(p_i)} \frac{PR(p_j)}{L(p_j)} \qquad (1)$$

Where $p_1, p_2, \ldots, p_n$ are the nodes in the network and N is the total number of nodes, $M(p_i)$ is the set of nodes that link to $p_i$, $L(p_j)$ is the sum of weights of out-going links on node $p_j$, $PR(p_i)$ is the probability that the random surfer is on node $p_i$, then same for $PR(p_j)$. $d$ is the damping factor which is the probability that a random surfer will follow one of the links on the present page. In this paper, the damping factor was set to 0.15 (to stress the equal chance of being cited), 0.5 (to indicate that scientific papers usually follow a short path of 2), or 0.85 (to stress the network topology) (Chen, Xie, Maslov, & Redner, 2007).

The weighted PageRank was proposed as

$$PR_w(p_i) = (1 - d) \times \frac{W(p_i)}{\sum_{k=1}^{N} W(p_k)} + d \sum_{p_j \in M(p_i)} \frac{PR_w(p_j)}{L(p_j)} \quad (2)$$

Where $W(p_i)$ is the weight assigned to node $p_i$, $\sum_{k=1}^{N} W(p_k)$ is the sum of the weights assigned to each node in the network. $PR\_w(p_i)$ is the weighted PageRank score for $p_i$ and same for $PR\_w(p_j)$. So in weighted PageRank formula, every node does not have an equal chance to get visited by the random surfer, while nodes with high weights will get high chance. The weight for each node can be the number of citations received by this node, the number of publications where this node acts as a first author, or the h-index of this node. In this paper, two weighted vectors were considered: the number of citations of the nodes and the number of publications of the nodes. It indicates that nodes with large number of citations or publications will have high probability to get visited by the random surfer.

The author citation network is a directed and weighted graph where nodes represent authors, edges represent citing relationships from author A to author B, and edge weights represent the number of times that author A cites author B. For each phase, papers without any reference were ignored. Four author citation networks were built: the 5395*5395 matrix in Phase 1, the 5978*5978 matrix in Phase 2, the 33483*33483 matrix in Phase 3, and the 61013*61013 matrix in Phase 4. The original PageRank (denoted PR) and weighted PageRank (denoted PR_c or PR_p where citation or publication, respectively, is the weighted vector) were calculated based on these author citation networks to see whether they can measure the popularity and prestige of authors. The original and weighted PageRank were calculated using a PageRank MATLAB program with damping factor of 0.15, 0.5, and 0.85.

For each phase, the top 20 and the top 100 highly cited authors were chosen to test the variations of their PageRanks and weighted PageRanks with their popularity ranks and prestige ranks, and to identify the correlation variation at different ranking levels (top 20 vs. top 100). Spearman's rank correlation test (two-tailed) was used to calculate the correlations among the ranks of these measures for the top 20 or top 100 highly cited authors respectively. Two indicators (the h-index and impact factor) were added to measure the 2001-2008 dataset. Principal Component Analysis (PCA) was conducted to detect relationships among these different measures. Finally, the coverage of IR prize winners captured by these measures was compared at the top 5, 10, 20, and 50 ranks.

## Results

**Top 20 Ranks**

The ranks of the top 20 authors measured by citation, PR ($d$), PR_c($d$), and PR_p($d$), where $d$ is 0.85, 0.5, or 0.15 are shown in Appendix. The dynamic changes of their ranks were tested in four different periods. The top 20 prestigious authors are the top 20 authors according to prestige rank, which was calculated based on how many times the author has been cited by highly cited papers (for more details, see Ding & Cronin, 2010 forthcoming). Figure 3 shows the variety of changes in the popularity rank, prestige rank, PageRank, and weighted PageRank in the four phases. For the top 20s, prestige rank differed dramatically from the popularity rank, PageRank, and weighted PageRank.

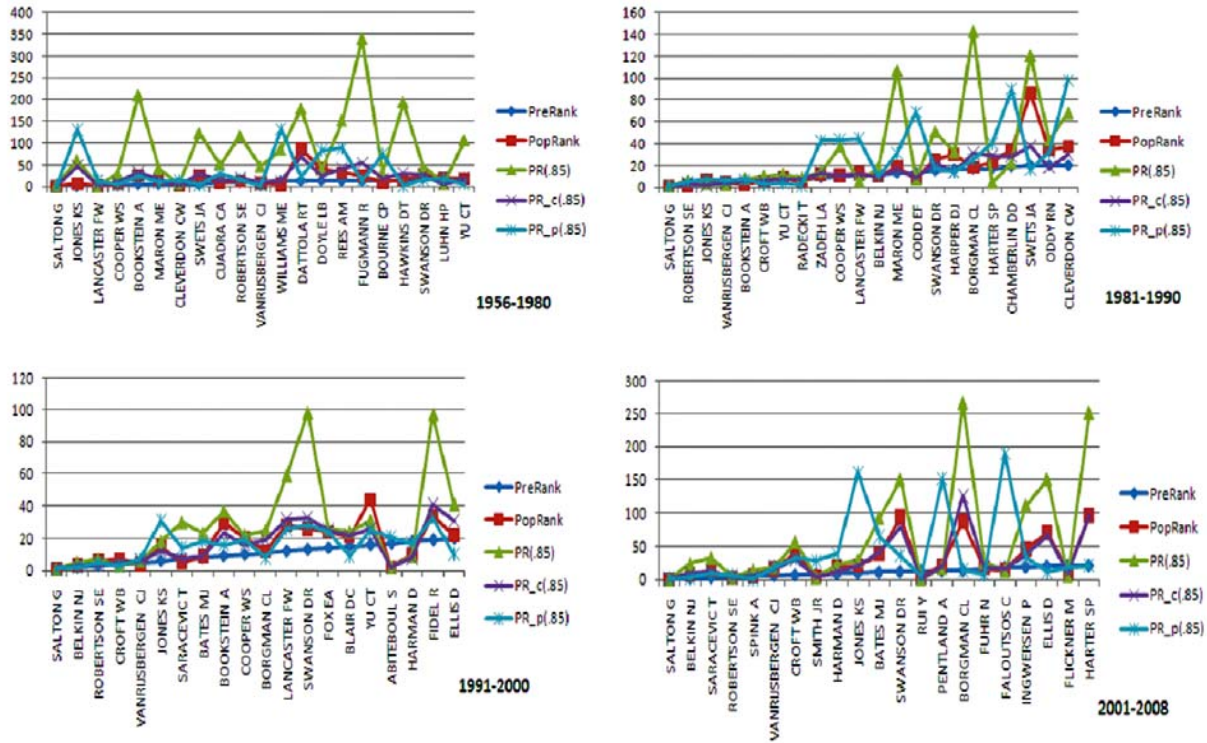

Figure 3. Ranking of the top 20 prestigious authors based on popularity rank, PageRank, and weighted PageRank in the four periods (PreRank represents prestige rank and PopRank represents popularity rank).

**Top 100 Ranks**

The Spearman correlation test (two-tailed) on the top 100 highly cited authors was calculated to identify correlations among these measures (see Table 1). Popularity rank and prestige rank were significantly correlated in Phase 1($r=0.548$, $p<0.01$), Phase 2 ($r=0.744$, $p>0.01$), Phase 3 ($r=0.662$, $p<0.01$), and Phase 4 ($r=0.490$, $p<0.01$). Table 2 summarizes the correlations for the top 100 highly cited authors. Figure 4 displays the scatter plots of the correlations. In general, popularity rank had a higher correlation with PR_c (mean $r=0.925$, $p<0.01$) than with PR (mean $r=0.702$, $p<0.01$) or PR_p (mean $r=0.485$, $p<0.01$), and had the highest correlation with PR_c(.15) (mean r=0.931, p<0.01). Prestige rank had a higher correlation with PR_c (mean r=0.616, p<0.01) than with PR (mean r=0.379, p<0.01) or PR_p (mean r=0.588, p<0.01), and had the highest correlation especially with PR_c(.85) in Phase 2 and PR_c(.85) in Phase 3.

Table 1. The Spearman correlation test of different ranks for top 100 highly cited authors

|  | PopRank | PreRank | PR (.85) | PR (.5) | PR (.15) | PR_c (.85) | PR_c (.5) | PR_c (.15) | PR_p (.85) | PR_p (.5) | PR_p (.15) |
|---|---|---|---|---|---|---|---|---|---|---|---|
| 1956-1980 ||||||||||||
| PopRank | 1 | | | | | | | | | | |
| PreRank | 0.548 | 1 | | | | | | | | | |
| PR(.85) | 0.524 | 0.201* | 1 | | | | | | | | |
| PR(.5) | 0.544 | 0.183* | 0.999 | 1 | | | | | | | |
| PR(.15) | 0.561 | 0.18* | 0.969 | 0.99 | 1 | | | | | | |
| PR_c(.85) | 0.764 | 0.447 | 0.833 | 0.811 | 0.783 | 1 | | | | | |
| PR_c(.5) | 0.816 | 0.486 | 0.796 | 0.785 | 0.766 | 0.989 | 1 | | | | |
| PR_c(.15) | 0.866 | 0.484 | 0.726 | 0.731 | 0.733 | 0.935 | 0.973 | 1 | | | |
| PR_p(.85) | 0.451 | 0.465 | 0.512 | 0.485 | 0.464 | 0.64 | 0.63 | 0.585 | 1 | | |
| PR_p(.5) | 0.448 | 0.486 | 0.44 | 0.417 | 0.402 | 0.579 | 0.577 | 0.544 | 0.987 | 1 | |
| PR_p(.15) | 0.454 | 0.496 | 0.409 | 0.39 | 0.379 | 0.552 | 0.557 | 0.535 | 0.974 | 0.996 | 1 |
| 1981-1990 ||||||||||||
| PopRank | 1 | | | | | | | | | | |
| PreRank | 0.744 | 1 | | | | | | | | | |
| PR(.85) | 0.774 | 0.601 | 1 | | | | | | | | |
| PR(.5) | 0.763 | 0.551 | 0.987 | 1 | | | | | | | |
| PR(.15) | 0.739 | 0.498 | 0.955 | 0.987 | 1 | | | | | | |
| PR_c(.85) | 0.927 | 0.807 | 0.846 | 0.795 | 0.736 | 1 | | | | | |
| PR_c(.5) | 0.963 | 0.798 | 0.83 | 0.793 | 0.743 | 0.99 | 1 | | | | |
| PR_c(.15) | 0.994 | 0.764 | 0.799 | 0.779 | 0.745 | 0.956 | 0.984 | 1 | | | |
| PR_p(.85) | 0.638 | 0.73 | 0.498 | 0.443 | 0.396 | 0.721 | 0.71 | 0.668 | 1 | | |
| PR_p(.5) | 0.616 | 0.71 | 0.465 | 0.414 | 0.371 | 0.689 | 0.683 | 0.645 | 0.995 | 1 | |
| PR_p(.15) | 0.601 | 0.695 | 0.442 | 0.393 | 0.351 | 0.668 | 0.664 | 0.628 | 0.986 | 0.995 | 1 |
| 1991-2000 ||||||||||||
| PopRank | 1 | | | | | | | | | | |
| PreRank | 0.662 | 1 | | | | | | | | | |
| PR(.85) | 0.688 | 0.335 | 1 | | | | | | | | |
| PR(.5) | 0.664 | 0.268 | 0.983 | 1 | | | | | | | |
| PR(.15) | 0.633 | 0.202** | 0.936 | 0.982 | 1 | | | | | | |
| PR_c(.85) | 0.9 | 0.7 | 0.766 | 0.691 | 0.6 | 1 | | | | | |
| PR_c(.5) | 0.952 | 0.695 | 0.76 | 0.702 | 0.629 | 0.988 | 1 | | | | |
| PR_c(.15) | 0.994 | 0.677 | 0.72 | 0.685 | 0.638 | 0.938 | 0.978 | 1 | | | |
| PR_p(.85) | 0.533 | 0.666 | 0.33 | 0.251 | 0.169* | 0.626 | 0.603 | 0.565 | 1 | | |
| PR_p(.5) | 0.506 | 0.629 | 0.283 | 0.213** | 0.141* | 0.57 | 0.554 | 0.53 | 0.992 | 1 | |
| PR_p(.15) | 0.489 | 0.616 | 0.259 | 0.192** | 0.123** | 0.546 | 0.531 | 0.512 | 0.983 | 0.997 | 1 |
| 2001-2008 ||||||||||||
| PopRank | 1 | | | | | | | | | | |
| PreRank | 0.49 | 1 | | | | | | | | | |
| PR(.85) | 0.855 | 0.299 | 1 | | | | | | | | |
| PR(.5) | 0.847 | 0.269 | 0.994 | 1 | | | | | | | |
| PR(.15) | 0.827 | 0.236** | 0.974 | 0.992 | 1 | | | | | | |
| PR_c(.85) | 0.947 | 0.514 | 0.873 | 0.84 | 0.793 | 1 | | | | | |
| PR_c(.5) | 0.977 | 0.514 | 0.877 | 0.853 | 0.816 | 0.991 | 1 | | | | |
| PR_c(.15) | 0.996 | 0.509 | 0.865 | 0.852 | 0.826 | 0.965 | 0.989 | 1 | | | |
| PR_p(.85) | 0.43 | 0.585 | 0.22** | 0.186** | 0.155* | 0.442 | 0.437 | 0.437 | 1 | | |
| PR_p(.5) | 0.337 | 0.497 | 0.135* | 0.109* | 0.088* | 0.33 | 0.329 | 0.338 | 0.982 | 1 | |
| PR_p(.15) | 0.312 | 0.479 | 0.11* | 0.086* | 0.068* | 0.301 | 0.302 | 0.313 | 0.974 | 0.998 | 1 |

Note: * means that they are not significantly correlated at the 0.05 confidence level; ** means that they are not significantly correlated at the 0.01 confidence level;

Table 2. Summary of the correlations for top 100 highly cited authors.

| Top 100 highly cited authors | 1956-1980 | 1981-1990 | 1991-2000 | 2001-2008 |
|---|---|---|---|---|
| popularity vs. prestige | 0.548 | 0.744 | 0.662 | 0.490 |

| popularity vs. original PR | 0.543 | 0.759 | 0.662 | 0.843 |
|---|---|---|---|---|
| popularity vs. weighted PR_c | 0.815 | 0.961 | 0.949 | 0.973 |
| popularity vs. weighted PR_p | 0.451 | 0.618 | 0.509 | 0.360 |
| highest correlation with popularity | PR_c(.15) (0.866) | PR_c(.15) (0.994) | PR_c(.15) (0.994) | PR_c(.15) (0.996) |
| prestige vs. original PR | 0.188* | 0.55 | 0.302 | 0.284 |
| prestige vs. weighted PR_c | 0.472 | 0.790 | 0.691 | 0.512 |
| prestige vs. weighted PR_p | 0.482 | 0.712 | 0.637 | 0.520 |
| highest correlation with prestige | PR_p(.15) (0.496) | PR_c(.85) (0.807) | PR_c(.85) (0.7) | PR_p(.85) (0.585) |

Note: * means that they are not significantly correlated at the 0.05 confidence level;

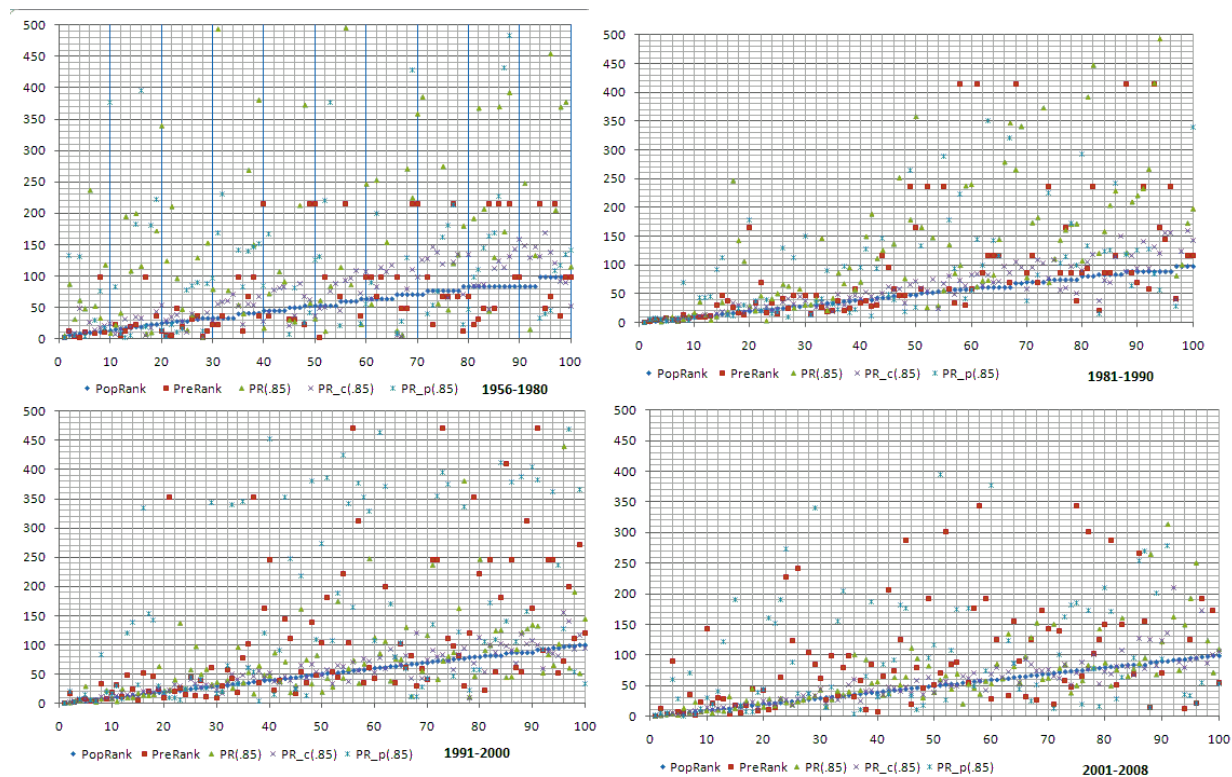

Figure 4. Scatter plots of popularity rank, prestige rank, original PageRank, and weighted PageRank. (X axis represents the authors (numbered according to the popularity rank). Y axis represents the ranks).

For different rank scales (top 20 vs. top 100), although the popularity rank and prestige rank differ for the top 20 highly cited authors, they were significantly correlated for the top 100 highly cited authors in these four phases. This result shows that the ranking scale matters: in our case, the discrepancy in the top 20 was larger than in the top 100. Bollen, Rodriguez, and Van de Sompel (2006) found similar results, in that the top 10 ranked journals based on the ISI IF and Y-factors significantly diverged, but significantly correlated when the scale increased. Chen, Xie, Maslov, and Redner (2007) also found a similar phenomenon in Physics, that is, only four papers were common in the top 10 highly ranked physics papers based on PageRank and citation counts, but they became significantly correlated when the scale increased.

**Correlation of various indicators**

Papers published in Phase 4 and their citations were selected to test the correlations among different indicators since 2001-2008 contained the largest number of papers and citations. Two new indicators were added to the above 11 indicators: the h-index rank and impact factor rank (IF rank). The h-index rank was calculated based on ranking the h-indexes of the top 100 highly cited authors from WOS. The IF rank was calculated by first adding the journal impact factor of the citing article to its citations, summing together the impact factors of citations for each of the top 100 highly cited authors, and then ranking these authors based on the summarized impact factors. The journal impact factor of the citing article is the corresponding journal impact factor based on the publication year of the citing article as journal impact factors vary yearly. Table 3 shows that prestige rank was highly correlated with PR_p(.85), which has publication counts as the weighted vector and stresses the author citation graph topology. The IF rank had a higher correlation with popularity rank than with prestige rank. The h-index rank was significantly correlated with popularity rank, prestiage rank, PR_c, PR_p(.85) or PR_p(.5) at a confidence level of 0.05.

Table 3. Correlations of 13 different indicators for the top 100 highly cited authors in 2001-2008.

| | PopRank | PreRank | PR(.85) | PR(.5) | PR(.15) | PR_c(.85) | PR_c(.5) | PR_c(.15) | PR_p(.85) | PR_p(.5) | PR_p(.15) | h-index Rank | IF Rank |
|---|---|---|---|---|---|---|---|---|---|---|---|---|---|
| PopRank | 1 | | | | | | | | | | | | |
| PreRank | 0.49 | 1 | | | | | | | | | | | |
| PR(.85) | 0.855 | 0.299 | 1 | | | | | | | | | | |
| PR(.5) | 0.847 | 0.269 | 0.994 | 1 | | | | | | | | | |
| PR(.15) | 0.827 | 0.236** | 0.974 | 0.992 | 1 | | | | | | | | |
| PR_c(.85) | 0.947 | 0.514 | 0.873 | 0.84 | 0.793 | 1 | | | | | | | |
| PR_c(.5) | 0.977 | 0.514 | 0.877 | 0.853 | 0.816 | 0.991 | 1 | | | | | | |
| PR_c(.15) | 0.996 | 0.509 | 0.865 | 0.852 | 0.826 | 0.965 | 0.989 | 1 | | | | | |
| PR_p(.85) | 0.43 | 0.585 | 0.22** | 0.186** | 0.155* | 0.442 | 0.437 | 0.437 | 1 | | | | |
| PR_p(.5) | 0.337 | 0.497 | 0.135* | 0.109* | 0.088* | 0.33 | 0.329 | 0.338 | 0.982 | 1 | | | |
| PR_p(.15) | 0.312 | 0.479 | 0.11* | 0.086* | 0.068* | 0.301 | 0.302 | 0.313 | 0.974 | 0.998 | 1 | | |
| h-index Rank | 0.167* | -0.065* | 0.188** | 0.192** | 0.188** | 0.144* | 0.153* | 0.157* | -0.183* | -0.184* | -0.192** | 1 | |
| IF Rank | 0.799 | 0.552 | 0.679 | 0.663 | 0.634 | 0.802 | 0.809 | 0.809 | 0.469 | 0.384 | 0.357 | 0.075* | 1 |

\* means that they are not significantly correlated at the 0.05 confidence level; ** means that they are not significantly correlated at the 0.01 confidence level;

Principal component analysis (PCA) was conducted for these 13 different measures. Four components were extracted that explained 87.84% of the total variance (Rotation method: Varimax with Kaiser Normalization). In Table 4, representative variables for each component were highlighted in bold if their loadings were more than 0.4 (Raubenheimer, 2004). Popularity rank and prestige rank belong to different components. Component 1 can be explained to represent the dimension of popularity highlighted by popularity rank, PR_c, and IF rank. Component 2 can be explained to represent the dimension of PageRank highlighted by PR and PR_c(.85). Component 3 can be explained to represent the dimension of weighted PageRank highlighted by PR_p. Component 4 can be explained to represent the dimension of prestige highlighted by prestige rank, h-index rank and IF rank. Therefore, the h-index rank and IF rank showed a significant level of prestige with very high absolute loadings in Component 4.

Table 4. PCA of 13 different indicators.

| | Component | | | |
|---|---|---|---|---|
| | 1 | 2 | 3 | 4 |
| PopRank | **.894** | .329 | .169 | -.083 |
| PreRank | .389 | .025 | .365 | **.535** |
| PR(.85) | .294 | **.944** | -.044 | -.016 |
| PR(.5) | .305 | **.947** | -.058 | -.043 |
| PR(.15) | .307 | **.932** | -.064 | -.063 |
| PR_c(.85) | **.848** | **.439** | .139 | .001 |
| PR_c(.5) | **.889** | .394 | .154 | -.032 |
| PR_c(.15) | **.898** | .352 | .168 | -.062 |
| PR_p(.85) | .247 | -.028 | **.887** | .179 |
| PR_p(.5) | .069 | -.041 | **.974** | .049 |
| PR_p(.15) | .038 | -.047 | **.972** | .065 |
| h-index Rank | .185 | .091 | -.119 | **-.849** |
| IF Rank | **.587** | -.005 | -.065 | **.420** |

**Evaluation**

Table 5 shows the number of IR prize winners contained in the top 5, 10, 20, and 50 ranks of these 13 different indicators, according to the 2001-2008 dataset. The IR prize contains the Gerard Salton Award and the Tony Kent Strix Award. Overall, prestige rank outperformed all the other measures at all different top rank categories. Thus, when capturing the prize winner is important, it is better to use prestige rank.

Table 5. IR prize winners captured by the 13 indicators (2001-2008).

| | Top@5 | Top@10 | Top@20 | Top@50 |
|---|---|---|---|---|
| PopRank | 2 | 2 | 5 | 8 |
| PreRank | 3 | 7 | 8 | 9 |
| PR(.85) | 2 | 2 | 4 | 7 |
| PR(.5) | 2 | 2 | 3 | 7 |
| PR(.15) | 1 | 2 | 3 | 7 |
| PR_c(.85) | 2 | 2 | 5 | 8 |
| PR_c(.5) | 2 | 2 | 6 | 8 |
| PR_c(.15) | 2 | 2 | 5 | 8 |
| PR_p(.85) | 1 | 3 | 5 | 6 |
| PR_p(.5) | 1 | 1 | 4 | 6 |
| PR_p(.15) | 1 | 1 | 2 | 6 |
| h-index Rank | 0 | 1 | 2 | 2 |
| IF Rank | 2 | 3 | 6 | 7 |

## Conclusion

This paper conducted a detailed analysis of popularity and prestige rankings based on data collected from WOS in IR for the period of 1956-2008. The whole 52-year period was divided into four phases. A weighted PageRank was proposed and calculated based on author citation networks. The proposed weighted PageRank contains different weighted vectors: the number of citations or the number of publications. Different damping factors were tested to stress the random citation (d=0.15), network topology (d=0.85), or short path of two for scientific papers (d=0.5). For the top 20 ranks, there were various discrepancies among prestige rank, popularity rank, PageRank, and weighted PageRank in different time periods. For the top 100, popularity rank and prestige rank were significantly correlated in all four phases.

Two measures (h-index rank and IF rank) were added to the above indicators to test the correlations in the 2001-2008 dataset. Four components were extracted by PCA representing: popularity, original PageRank, weighted PageRank and prestige. Popularity rank and prestige rank were separated to different components which indicated the difference of their measuring. The number of prize winners in the top 5, 10, 20, and 50 ranks of these 13 measures were also tested in 2001-2008. Overall, prestige rank outperformed all other indicators at all different top rank categories.

In bibliometrics, research impact is measured by different indicators using various counting methods on different datasets. Thus there are three important methodological components to the evaluation of scholarly impact: indicators, counting methods, and datasets. Most indicators are one-dimensional measures of either publications or citations. The recently developed h-index combines publications with citations. A future trend will be to generate indicators that measure multi-dimensional units (e.g., publications, citations, and co-authorship collaborations). Counting methods are generally limited to the simple counting of citations, and it is necessary to differentiate citations coming from Paper "The-Best" or Paper "The Worst." Adding weights to citations is therefore important. Finally, citing and co-authoring create relationships to link data from different datasets and form scholarly networks. Previous work focuses mainly on homogenous networks, where nodes in the network belong to the same data type. Since different types of data can be linked, it is meaningful to create new measures or algorithms to evaluate heterogeneous networks.

**Acknowledgement**

The author would like to thank two anonymous reviewers who help to improve the quality of this paper dramatically.

Footnotes

[1] http://www.etymonline.com/index.php?term=popular

[2] http://www.etymonline.com/index.php?search=prestige, http://dictionary.reference.com/browse/prestige